\documentclass[conference]{IEEEtran}
\IEEEoverridecommandlockouts

\usepackage{array}    
\newcolumntype{C}[1]{>{\centering\arraybackslash}p{#1}}
\usepackage{amsmath,amssymb,amsfonts}
\usepackage{algorithmic}
\usepackage{graphicx}
\usepackage{textcomp}
\usepackage{xcolor}
\usepackage[strings]{underscore}
\usepackage{url}

\usepackage{rotating}
\usepackage{multirow} 
\usepackage{makecell}
\usepackage{diagbox}

\usepackage{listings}
\usepackage{color}

\usepackage[capitalise,noabbrev,nameinlink]{cleveref}

\usepackage[style=ieee,backend=biber]{biblatex}  
\def\BibTeX{{\rm B\kern-.05em{\sc i\kern-.025em b}\kern-.08em
    T\kern-.1667em\lower.7ex\hbox{E}\kern-.125emX}}
\begin{document}

\title{SoK: Cross-Chain Bridging Architectural Design Flaws and Mitigations  \\

}

\author{\IEEEauthorblockN{1\textsuperscript{st} Jakob Svennevik Notland}
\IEEEauthorblockA{\textit{Department of Computer Science} \\
\textit{Norwegian University of Science and Technology}\\
Trondheim, Norway \\
jakob.notland@ntnu.no}
\and
\IEEEauthorblockN{2\textsuperscript{nd} Jingyue Li}
\IEEEauthorblockA{\textit{Department of Computer Science} \\
\textit{Norwegian University of Science and Technology}\\
Trondheim, Norway \\
jingyue.li@ntnu.no}
\and
\IEEEauthorblockN{3\textsuperscript{rd} Mariusz Nowostawski}
\IEEEauthorblockA{\textit{Department of Computer Science} \\
\textit{Norwegian University of Science and Technology}\\
Gjøvik, Norway \\
mariusz.nowostawski@ntnu.no}
\and
\IEEEauthorblockN{4\textsuperscript{th} Peter Halland Haro}
\IEEEauthorblockA{
\textit{Sintef Nord}\\
Tromsø, Norway \\
peter.haro@sintef.no}
}

\maketitle

\begin{abstract}
Cross-chain bridges are solutions that enable interoperability between heterogeneous blockchains. In contrast to the underlying blockchains, the bridges often provide inferior security guarantees and have been targets of hacks causing damage in the range of 1.5 to 2 billion USD in 2022.
The current state of bridge architectures is that they are ambiguous, and there is next to no notion of how different architectures and their components are related to different vulnerabilities. Throughout this study, we have analysed 60 different bridges and 34 bridge exploits in the last three years (2021-2023).
Our analyses identified 13 architectural components of the bridges. We linked the components to eight types of vulnerabilities, also called design flaws. We identified prevention measures and proposed  11 impact reduction measures based on the existing and possible countermeasures to address the imminent exploits of the design flaws. The results are meant to be used as guidelines for designing and implementing secure cross-chain bridge architectures, preventing design flaws, and mitigating the negative impacts of exploits.
\end{abstract}

\begin{IEEEkeywords}
Blockchain, cross-chain, DeFi, bridges, architectures, design flaws, mitigation.
\end{IEEEkeywords}

\section{Introduction}
\label{sec:introduction}

Blockchain technologies have been envisioned to break down the barriers of Web2.0's data silos~\cite{tapscott2016blockchain}.
However, blockchains have created new silos in different forms, programming languages, consensus algorithms, and structures.
Essentially, a heterogeneous environment of incompatible blockchains has emerged.
Furthermore, to address scalability challenges, the complexity is growing through developing second-layer solutions with sidechains, multichain ecosystems, payment channels, and relayers~\cite{10.1145/3471140,Bridges-Between-Islands-Cross-Chain-Technology}. The complex landscape motivates research on new solutions for secure and efficient blockchain interoperability.


Cross-chain bridges have been developed as interoperability solutions and uncover new security challenges. Bridges had a turbulent onset, reporting losses between 1.5 and 2 billion USD due to exploits of bridge vulnerabilities in 2022~\cite{beosinReport,chainalysis,certik}. The disparity between the reports may stem from combinations of diverse factors, the numerous incidents, and the ambiguous boundaries defining what qualifies as a bridge exploit.

Li et al.~present a broad approach when conducting a security analysis of decentralised finance (DeFi) to address the challenges. Their results show different vulnerabilities, attacks, and optimisations. The analysis includes bridges in general and other aspects in DeFi, such as decentralised exchanges (DEXes)~\cite{9881607}. Lee et al.~present a high-level overview of bridge architecture and suggest four main categories of attacks~\cite{10174993}. Zhang et al.~propose Xscope: a tool to detect cross-chain bridge attacks~\cite{10.1145/3551349.3559520}. They distinguish between three high-level types of bugs that lead to exploits.
The common denominator for related work is ambiguous bridge architectures, arguing the need for a fine-grained understanding of the components of the bridge architecture to identify the attack surfaces and design flaws and to propose precise attack countermeasures and exploits' impact mitigation. Thus, our study focuses on answering the following three research questions (RQs):

\begin{itemize}
  \item RQ1: What are the components of various cross-chain bridge architectures?
  \item RQ2: What are the design flaws of bridges, and what are the countermeasures against the design flaw exploitations? 
  \item RQ3: Which methods exist for reducing the impact of bridge design flaw exploits?
\end{itemize}

We see RQ1 as a necessary first step in this study to give people instructions and guidelines for making a bridge. This can help find which components to use depending on the architecture. RQ2 addresses the connection between the said components and bridge vulnerabilities and seeks to provide prevention measures. RQ3 addresses exploits where the breach has already happened. In these cases, we want to advise on performing effective impact-reduction measures.

We aim to gather as many bridge attacks as possible from 2021 and throughout the first half of 2023. The primary sources were Rekt\footnote{\url{https://rekt.news}}, Beosin\footnote{\url{https://beosin.com/}}, and Halborn\footnote{\url{https://halborn.com/}}. Rekt is an anonymous publification portal and provides the most comprehensive overview of Decentralised Finance (DeFi) exploits, often with technical-level descriptions of attacks and references to official reports. Beosin and Halborn are blockchain security companies providing services like smart contract audits and anomaly detection for DeFi projects. They have written several reports on individual events and summaries like yearly reviews.

Furthermore, we screened several secondary archives looking for additional incidents, including SunWeb3Sec\footnote{\url{https://github.com/SunWeb3Sec/DeFiHackLabs}}, Certik\footnote{\url{https://certik.com/}}, and Immunefi\footnote{\url{https://immunefi.com/}}. These archives did not reveal additional bridge exploits and were often linked to the same reports as the primary sources. Therefore, the incident reports were not systematically sampled from these archives.


To complement the search in the grey literature, we also searched scientific articles through Scopus\footnote{\url{https://scopus.com}} and the Web of Science\footnote{\url{https://webofscience.com/}}. However, except  ~\cite{9881607,10.1145/3551349.3559520,10174993}, most scientific articles address DeFi in general or lean towards other aspects, such as decentralised exchanges (DEXes) and maximum extractable value (MEV), which are not the focus of this study. From the grey and scientific literature, we assessed 60 bridges and 34 bridge exploits in 2021-2023.




To answer RQ1, we identified the architectural components of the bridges and clustered the components into different categories. To answer RQ2, we identified the location of vulnerabilities at the component level and categorised the vulnerabilities as design flaws. We also explored the possible vulnerability prevention and categorised them. To answer RQ3, we summarised the exploitation impact proposals. 

We used data triangulation~\cite{oates2006researching} throughout the analysis to reduce the bias of relying on information about a single source per exploit.
We also wanted to ensure that we include a statement from the responsible team and a third party to cross-validate our findings and detect potential conflicts or discrepancies. For instance, pNetwork claims that their actions saved pGala from being exploited~\cite{pGalaAttack}. On the other hand, Gala filed a lawsuit against pNetwork, claiming that pNetwork was enabling the vulnerability by leaking a private key~\cite{galaLawSuit}. In cases of uncertainty, we choose to either label the root cause as unknown or provide references to both perspectives.

The contributions of the study are:
\begin{itemize}

\item We categorised 13 components essential for composing bridges and linked them to different bridge architectures (RQ1). 

\item We identified eight critical vulnerabilities, tied them to the bridge components and suggested prevention measures (RQ2). 

\item We summarised 11 impact reduction measures and elaborated on how they could be effective depending on the exploited component (RQ3).
\end{itemize}

The structure of this article is as follows:~\cref{sec:background} introduces bridge designs and use cases for interoperability.~\cref{sec:researchImplementation} describes the research implementation of this study. \cref{sec:components} presents bridge architectures and related components. \cref{sec:bridgeAttacks} shows bridge attacks and how they relate to different components. \cref{sec:impactReduction} suggests measures for impact reduction. The findings are discussed in~\cref{sec:discussion}, and we conclude our work in~\cref{sec:conclusion}.

\section{Blockchain bridges} \label{sec:background}

Blockchain bridges are infrastructures that enable interactions between two or more blockchains. These interactions include transferring fungible or non-fungible assets and transferring messages, including: data, function calls, and state. Bridges are emerging concepts, and many different specifications have been suggested and applied~\cite{10.1145/3471140,10174993,ethereumBridgeDefinitions,Bridges-Between-Islands-Cross-Chain-Technology}. The first formalization of bridge mechanics was introduced with the definition of sidechains in 2014~\cite{back2014enabling}. Sidechain-based bridges lock assets on a parent chain before unlocking them on a sidechain. Cryptographic proofs for each cross-chain transaction and simple payment verification (SPV) can ensure the correctness of transfers. These techniques are still popular in contemporary bridge implementations. Furthermore, Herlihy introduced cross-chain-atomic-swaps in 2018~\cite{10.1145/3212734.3212736}. Even though this technique enables trustless exchange of assets across chains, atomic swaps do not scale well and cannot facilitate bridges alone. That is because individual ask and bid orders might have to be published as transactions and require an exact match before trading can commence. However, some bridge implementations combine liquidity pools with atomic swaps to ensure the atomicity of cross-chain transfers~\cite{ethereumBridgeDefinitions}. So far, there is a lack of systematisation of knowledge regarding different types of bridge architectures, their components, and the cross-cutting relationship between the components and the architectures.

\section{Research implementation}
\label{sec:researchImplementation}

\begin{figure}
  \centering
  \includegraphics[width=1\columnwidth]{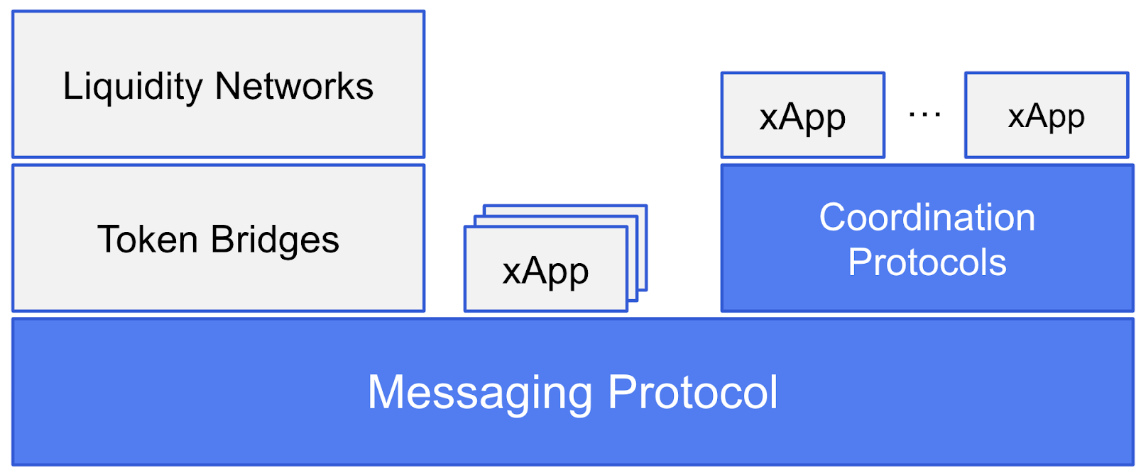}
  \caption{Layers of Crosschain Communication Protocols, retrieved from~\cite{crosschainriskframework}.}
  \label{fig:protocolstack}
\end{figure}

To research bridge architectures and their exploits and design flaws, we chose to use the cross-chain communication protocols stack proposed in the Cross-Chain Risk Framework~\cite{crosschainriskframework}, shown in~\cref{fig:protocolstack}, as the starting point. The framework has been developed to identify and address the risks of cross-chain bridge architecture, which is why it is intuitive for our purpose of making a systemization on bridge architectures and corresponding attacks. This is one of the ways that we differentiate ourselves from related work~\cite{9881607,10174993,10.1145/3551349.3559520}, which do not consider the nuances of different bridge architectures and their correspondence to design flaws and prevention measures.

The cross-chain communication protocol stack concerns two main aspects: \textit{the messaging protocols} that enable interactions between the chains and \textit{the different types of operation}, such as liquidity networks, token bridges, and coordination protocols. We rely on these two aspects to characterise the diverse bridge implementations. 

The Cross-Chain Risk Framework is an overarching representation of four main messaging protocols:

\begin{itemize}
  \item \textbf{State validating protocols:} The trust assumptions from underlying networks guarantee the state and state changes without compromise. This is typical for Layer~1 to Layer~2 bridges, where Layer~1 acts as the single source of truth.
  \item \textbf{Consensus verifying protocols:} The receiving network verifies that the transaction on the source chain has been finalised.
  \item \textbf{Third-party attestation protocols:} A trusted entity, a single validator, a consortium, or a separate blockchain validates transactions.
  \item \textbf{Optimistic protocols:} A single honest validator can detect and report fraudulent transactions to invalidate them if submitted within the time of a fraud window.
\end{itemize}

Furthermore, the Cross-Chain Risk Framework includes three bridge types:

\begin{itemize}
  \item \textbf{Liquidity networks:} Tokens are held in liquidity pools. A user can deposit tokens into the source chain pool and receive the equivalent amount from the liquidity pool on the destination chain.
  \item \textbf{Token bridges:} Tokens are locked or burned on the source chain before being unlocked or minted on the destination chain. This involves synthetic tokens representing the owner's right to redeem the native token.
  \item \textbf{Coordination protocols:} Functionality across different chains can be combined to create complex functionality. This involves cross-chain data sharing, function calls, and state synchronisation.
\end{itemize}

\begin{figure}
  \centering
  \includegraphics[width=1\columnwidth]{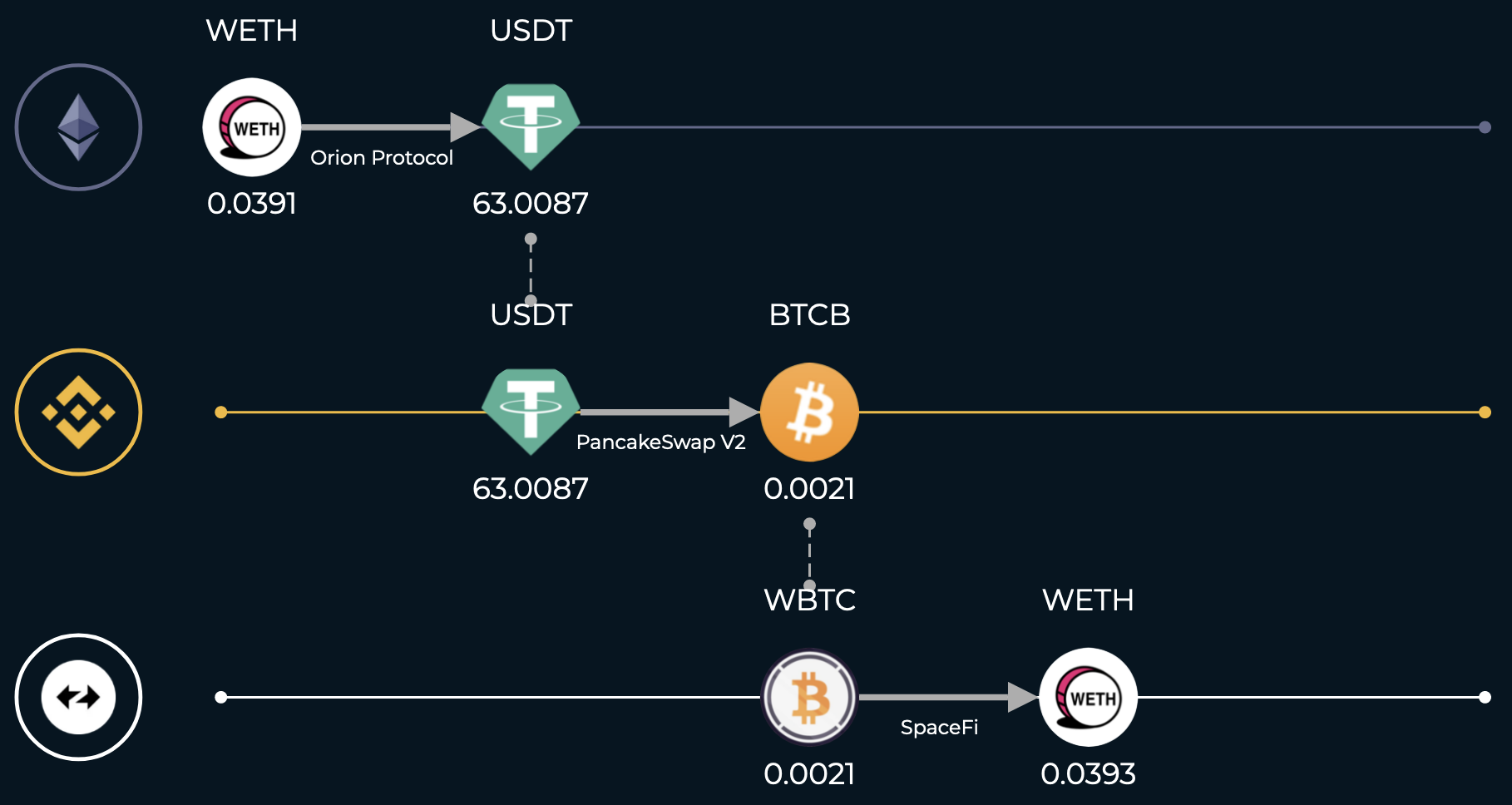}
  \caption{Cross-chain aggregated path from Odos~\cite{odos}. Path finding and single-chain swaps are considered out of scope, while the cross-chain transfers of USDT and BTCB to WBTC are in scope.}
  \label{fig:aggregator}
\end{figure}

Bridges are complex and often implemented through many different architectures. In addition to cross-chain transfers, some bridges perform swap operations before and after the bridging. Additionally, there exist aggregators that find the optimal path for a transfer through a combination of various swaps and bridges. We considered these aspects out of scope in this study as we wanted to focus on individual cross-chain interactions. 
\Cref{fig:aggregator} shows an example aggregated path from Ethereum to zkSync through the Binance chain. The cross-chain transfer of USDT from ETH to the Binance chain and BTCB from the Binance chain to WBTC on zkSync are considered in scope. On the other hand, path finding and on-chain swaps are considered out of scope.

\section{RQ1: Bridge components} 
\label{sec:components}

This section presents the bridge components discovered and their relation to the different bridge architectures. We first relied on the cross-chain risk framework to identify the main architectures composed of messaging protocol and bridge type. Furthermore, we identified bridge components by determining which components were targeted or could have been exposed in the bridge attack incidents. We commenced by connecting the components to the related architectures.

~\cref{tab:componentsArchitectures} is an abstraction of the layers of the Cross-Chain Communication Protocol and shows the low-level correlation between the 12 architectures that can be composed by the Cross-Chain Communication Protocol stack and the 13 components we have identified. The table also indicates whether the components are optional or mandatory depending on the chosen architecture. Various implementations of the components in the Cross-Chain Communication Protocols and their combinations lead to different concrete bridge architectures. The 12 different architectures are shown in each column as combinations of the communication protocol and bridge type, based on the stack shown in~\cref{fig:protocolstack}. For instance, a state-validating liquidity network (first architecture column) inherits the trust assumption from the underlying blockchain. It requires all the following components: Relayer, attester, deployer, and liquidity pool. The price calculation component is optional because several liquidity networks operate with source-chain pools that can only bridge to destination-chain pools with the same asset. We refer to these as 1:1 liquidity networks. Other liquidity networks operate as direct cross-chain exchanges and require price calculation. Bridge application frontend, backend, and DNS are also optional because, technically, a user could act as the relayer and make direct remote process calls against the blockchain to use the bridge service. However, it is more user-friendly if there is a hosted application. We describe each component in subsequent subsections.

\begin{table*}[!h!!t]
  \centering
  \caption{Overview of components that are included for different bridge architectures. An 'X' indicates that the component is required to build a given architecture. Furthermore, 'O' indicates that the component is optional for the given architecture. Abbreviations: LN - Liquidity Network, TB - Token Bridge, and CP - Coordination Protocol.~\textsuperscript{1}: Properties inherited from the underlying blockchain.~\textsuperscript{2} Token bridges can use 'burn-and-mint' techniques to omit the custodian.~\cite{burn-and-mint,circle-cross-chain}}
  \label{tab:componentsArchitectures}
  \begin{tabular}{|c|C{0.65cm}|C{0.65cm}|C{0.65cm}|C{0.65cm}|C{0.65cm}|C{0.65cm}|C{0.65cm}|C{0.65cm}|C{0.65cm}|C{0.65cm}|C{0.65cm}|C{0.65cm}|}
  \hline
  \multirow{2}{*}{\diagbox{Components}{Architectures}}
 & \multicolumn{3}{c|}{State validating} & \multicolumn{3}{c|}{Consensus verifying} & \multicolumn{3}{c|}{Third-party attestation} & \multicolumn{3}{c|}{Optimistic} \\
  & {LN} & {TB} & {CP} & {LN} & {TB} & {CP} & {LN} & {TB} & {CP} & {LN} & {TB} & {CP} \\
  \hline
  Trust assumption & \multicolumn{3}{c|}{Inherit~\textsuperscript{1}} & \multicolumn{3}{c|}{Inherit~\textsuperscript{1}} & \multicolumn{3}{c|}{m-n} & \multicolumn{3}{c|}{1-n} \\
  \hline
  Relayer & X & X & X & X & X & X & X & X & X & X & X & X \\
  \hline
  Attester & X~\textsuperscript{1} & X~\textsuperscript{1} & X~\textsuperscript{1} & X~\textsuperscript{1} & X~\textsuperscript{1} & X~\textsuperscript{1} & X & X & X & O & O & O \\
  \hline
  Custodian & & O~\textsuperscript{2} & & & O~\textsuperscript{2} & & & O~\textsuperscript{2} & & & O~\textsuperscript{2} & \\
  \hline
  Token interface & & O & & & O & & & O & & & O & \\
  \hline
  Debt issuer & & X & & & X & & & X & & & X & \\
  \hline
  Deployer & X & X & X & X & X & X & X & X & X & X & X & X \\
  \hline
  Watcher & & & & & & & & & & X & X & X \\
  \hline
  Liquidity pool & X & & & X & & & X & & & X & & \\
  \hline
  Price calculation & O & & & O & & & O & & & O & & \\
  \hline
  Bridge application frontend & O & O & O & O & O & O & O & O & O & O & O & O \\
  \hline
  Bridge application backend & O & O & O & O & O & O & X & X & X & O & O & O \\
  \hline
  Bridge application DNS & O & O & O & O & O & O & X & X & X & O & O & O \\
  \hline
  \end{tabular}
\end{table*}

\subsection{Trust assumption}

The trust assumption indicates the number of trusted entities required to secure the bridges. State validating and consensus verifying bridges inherit this attribute from the underlying blockchain, and the underlying protocol must enable the capabilities. Third-party attestation protocols could use a single-entity, multi-signature scheme or an intermediate blockchain. The strength of this trust assumption varies greatly depending on the implementation. Additionally, the optimistic bridges require only one honest party to prevent fraudulent transactions.

State validating protocols are only known to serve bridges for layer one to layer two transactions. As long as this is the case, these bridges resist reorg attacks. Blockchain reorg can happen if the blocks in one of the chains suspect a bridge is rewritten because of a rollback, conflicting blocks or 51\% attack. State validating protocols prevent this because the design relies on the layer one chain as a single source of truth, and if that chain reorgs, so will the layer two blockchain.

\subsection{Relayer}

The relayer is an entity making calls to execute cross-chain operations on the respective blockchains. These operations may or may not be privileged. For instance, some bridges have public functions that anyone can call if they provide a valid signature from attesters. Optimistic blockchains often allow anyone to become a relayer if they deposit collateral. This collateral is used to reward watchers submitting fraud proofs.

\subsection{Attester}

Most bridges require privileged attesters to execute bridge calls. For state validating and consensus verifying bridges, the consensus nodes in the underlying blockchain act as attesters. A third-party attestation protocol defines the attesters separately. Optimistic bridges do not require attesters because the trust falls back on the watchers. Watchers will be introduced in~\cref{sec:watchers}.

\subsection{Custodian}

The custodian is a component used by token bridges and serves as a vault to lock assets on the source chain before the debt issuer mints synthetic assets on another chain. This component can be seen as a prime target for adversaries as it often holds the value of several millions of dollars.

\subsection{Token interface}

Token interfaces enable the conversion of assets to synthetic assets. Token bridges often wrap native assets on the source chain, commencing the cross-chain transfer.

\subsection{Debt issuer}

The debt issuer is responsible for minting synthetic assets on the destination chain. This operation can commence once a deposit has been confirmed on the source chain. Debt issuers are also responsible for burning tokens on one chain before they can be unlocked or minted on another chain.

\subsection{Deployer}
The deployer has privileges to upgrade bridge contracts. This can, for instance, be used to deploy a new debt issuer with a different owner or set a new valid deposit address.

\subsection{Watchers} \label{sec:watchers}

Watchers enable optimistic bridges and can report adversarial activities by submitting fraud-proof within the timespan of the fraud window. Watchers are usually given a reward upon submission of valid fraud-proof. 

\subsection{Liquidity pool}

Liquidity pools enable cross-chain transfers as long as sufficient assets are available in the pools. A user deposits assets into a pool on the source chain before the equivalent amount is released on the destination chain. Furthermore, liquidity providers are incentivized to store assets in pools by gaining rewards. Liquidity networks typically rely on atomic swaps to enable cross-chain transfers~\cite{ethereumBridgeDefinitions}.

\subsection{Price calculation}

Price calculation is required for liquidity networks that perform direct cross-chain exchange of assets. This component will calculate the exchange rate based on the liquidity residing in the pools, similar to intrachain swaps like Uniswap\footnote{\url{https://uniswap.org}}.

\subsection{Bridge application frontend, backend and DNS}
Bridges are made available to users by a bridge application frontend that interacts with a bridge application backend to compose and forward the transactions to the bridge contracts. The Bridge application DNS maps the bridge application IP address to the corresponding domain. We include DNS as a component because we have seen it being hijacked in the Celer bridge through Border Gateway Protocol (BGP)~\cite{celerBGP}. However, we consider BGP out of scope because bridge developers cannot access it; thus, there are no corresponding examples of DNS exploits. We still include DNS because it can be accessed in other ways, such as through a cloud account.

\begin{figure}
    \centering
    \includegraphics[width=1\linewidth]{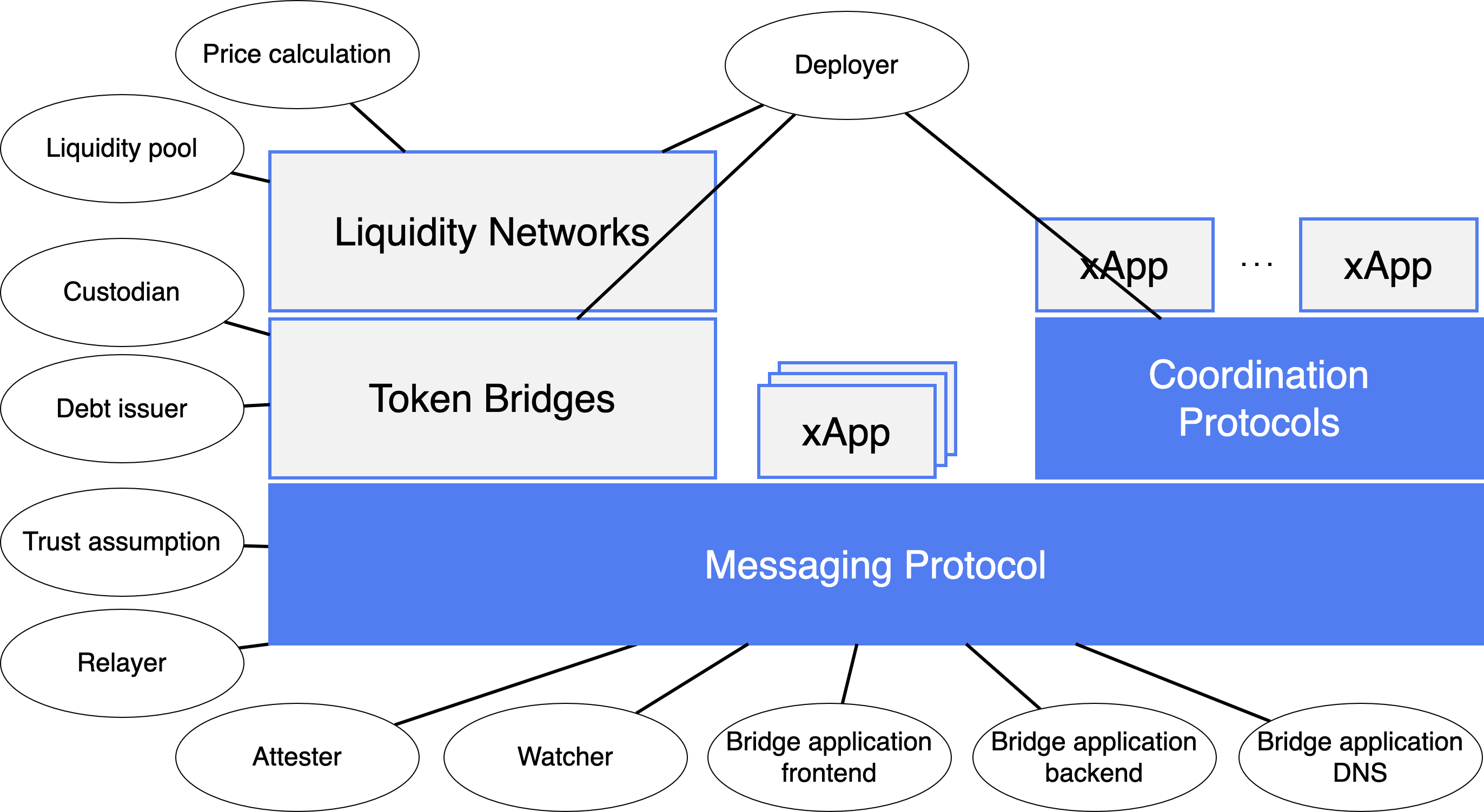}
    \caption{Shows the relation between the Cross-Chain Communication Protocol and components identified in the study.}
    \label{fig:componentsArchitectures}
\end{figure}

To summarize the findings of RQ1,~\cref{fig:componentsArchitectures} shows the connection between the components and the Cross-Chain Communication Protocol stack.


\section{RQ2: Bridge attacks}
\label{sec:bridgeAttacks}

\begin{figure}
    \centering
    \includegraphics[width=1\linewidth]{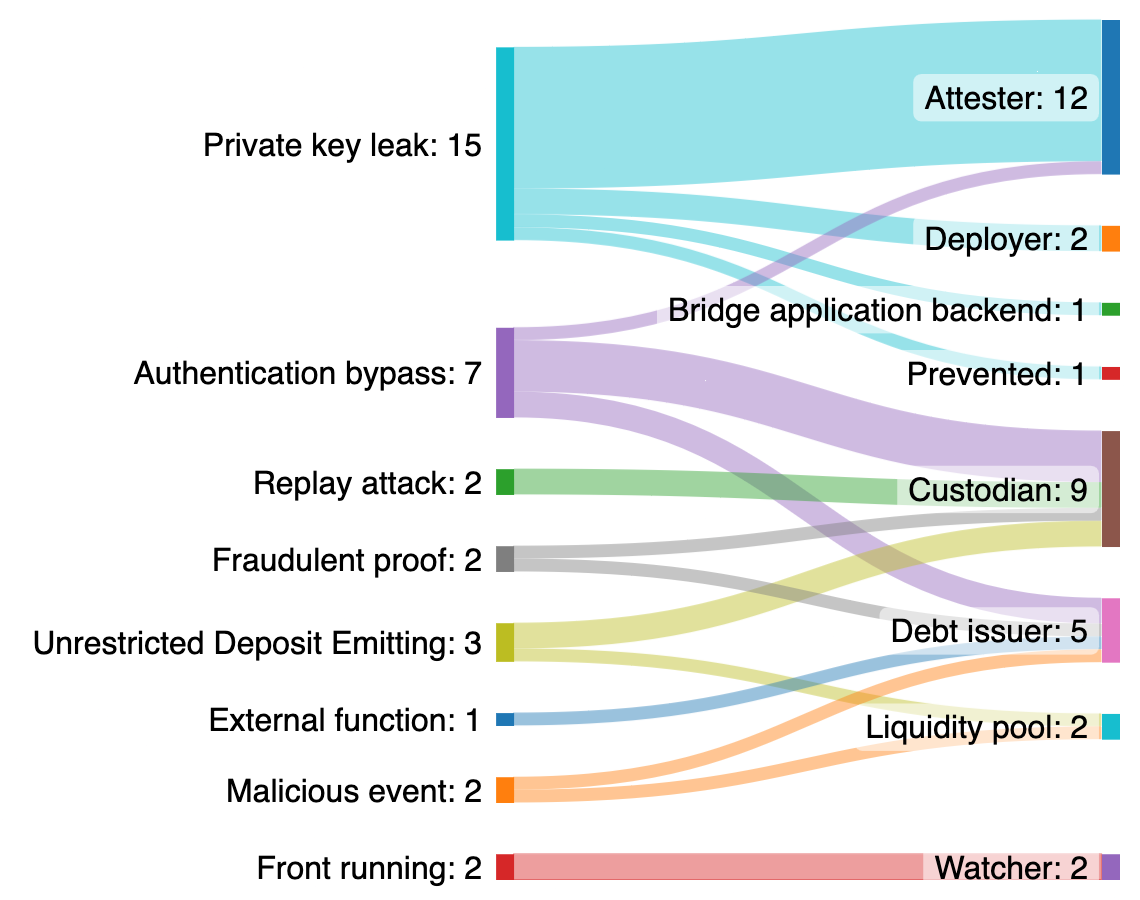}
    \caption{The Sankey diagram shows the exploited vulnerabilities and related components.}
    \label{fig:sankey}
\end{figure}

We sorted all the inspected bridges according to the cross-chain communication protocol stack to understand the causal connections between architectures and exploits. To summarize the findings, we show a Sankey diagram in~\cref{fig:sankey} to present the exploits assessed in this study, including the vulnerabilities and the target vulnerable components.

\begin{figure*}[h!!t!]
  \centering
  \includegraphics[width=0.85\textwidth]{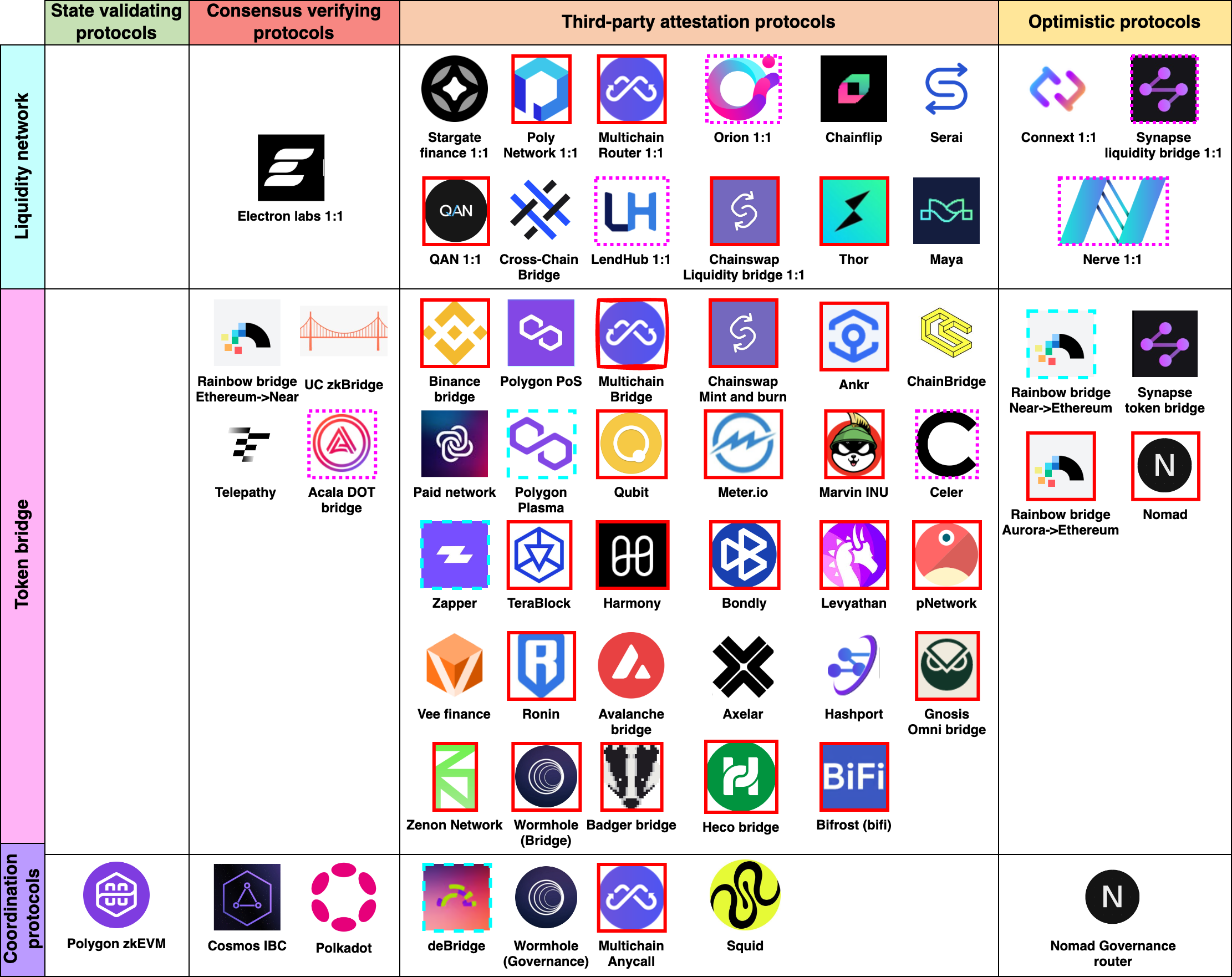}
  \caption{Categorisation of different bridges and their status of being exploited. Three kinds of borders indicate whether the bridges have been exploited: 1) Bridges with a solid red frame have been exploited. 2) Bridges with a dashed cyan frame had known vulnerabilities that were fixed before being exploited. 3) Bridge exploits with a pink dotted frame have are out of scope. 
  }
  \label{fig:solutionsCategories}

\vskip 1cm

    \begin{minipage}{0.54\textwidth}
        \centering
        \includegraphics[width=\textwidth]{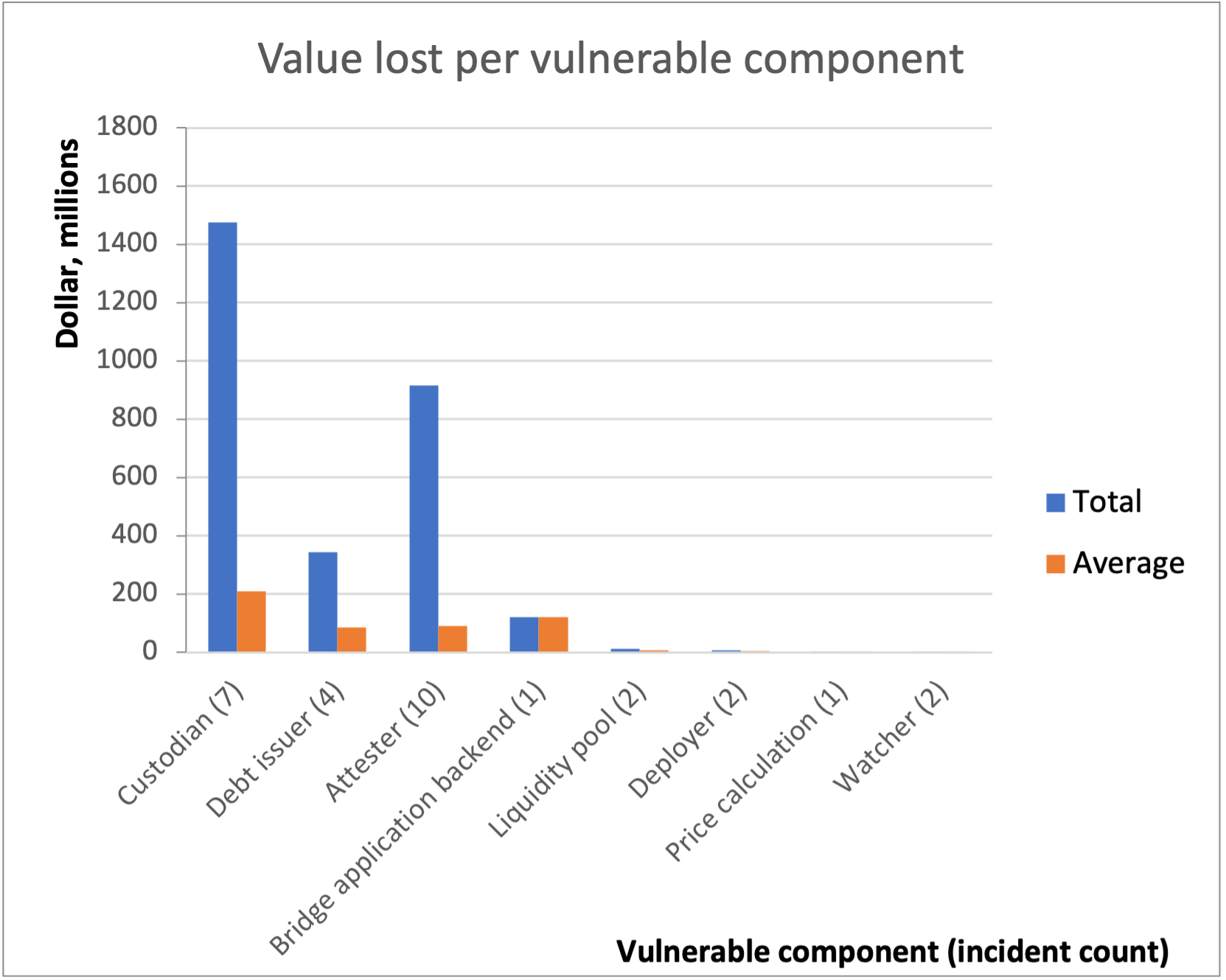} 
        \caption{Average and total value lost per vulnerale component.}
        \label{fig:image1}
    \end{minipage}\hfill
    \begin{minipage}{0.453\textwidth}
        \centering
        \includegraphics[width=\textwidth]{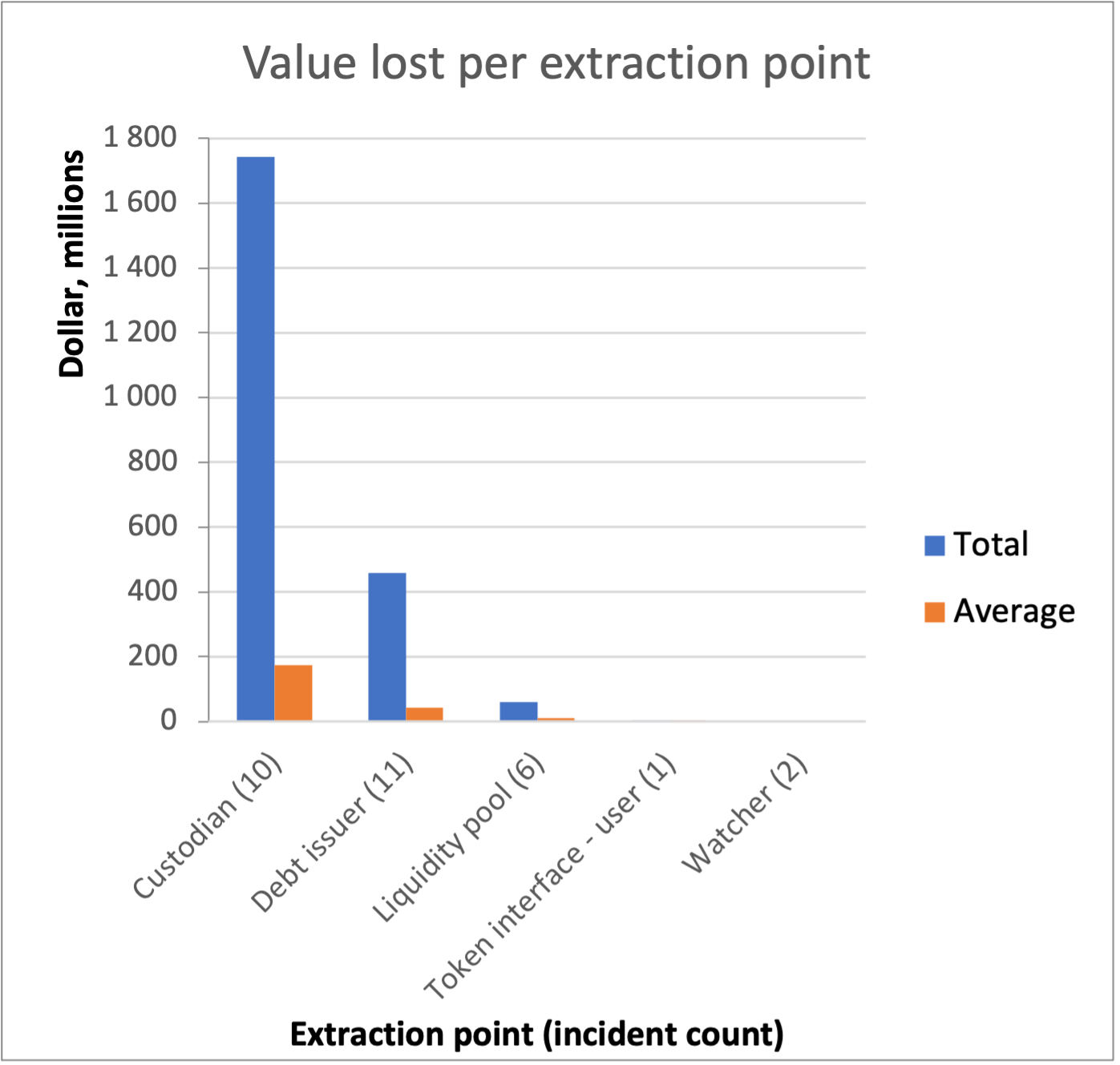} 
        \caption{Average and total value lost per extraction point.}
        \label{fig:image2}
    \end{minipage}
\end{figure*}

We further categories the inspected bridges as shown in~\cref{fig:solutionsCategories}, where bridges are arranged by communication protocol and the bridge type. The grid comprises the communication protocols as columns and bridge types as rows. The grid shows all the bridges considered during this study but is not necessarily an exhaustive collection of bridges. Bridges are marked with a red border to indicate an effective attack. The cyan dashed border indicates a known bug that was not exploited. At last, the pink dotted border indicates that the ecosystem has been targeted. Still, these vulnerabilities were out of scope, mainly related to inter-chain swaps.

~\Cref{fig:image1} shows which vulnerable components have been exploited to gain the most value, which is the custodian.~\Cref{fig:image2} shows the extraction point where the assets have been drained where the custodian is also the primary extraction point. The extraction point is either the vulnerable component or, in most cases, the accessory component whenever applicable in~\cref{tab:bridgeExploitTarget}. For instance, private key leak exploits mainly target the attester component. However, the assets reside in other components, such as the custodian. In that case, the custodian becomes the extraction point even though it is not vulnerable.

Our approach has been to search for bridges that have been exploited. Therefore, the representation is most complete for those marked by red frames. 
Some bridges have likely not been included in this study because there has not been an effective exploit of those bridges. 

We include two types of liquidity networks in this study as depicted in the~\cref{fig:solutionsCategories}.
The most common implementations are liquidity networks where the bridges serve the same asset on both the source and destination chain, such as Poly Network~\cite{polyNetwork}, and these assets are only transferred in a {\tt 1:1} fashion between chains.
This requires no price calculation of the asset transfer. In contrast, cross-chain exchanges, such as ThorChain~\cite{thorChain}, using liquidity networks for direct cross-chain exchange have additional complexity.

The attacks were also sorted according to the different types of vulnerabilities and targeted components, as shown in~\cref{tab:bridgeExploitTarget}. The table presents an overview and indicates the most common vulnerabilities, relevant components and root causes for cross-chain bridge exploits.


The final results are shown in~\cref{tab:vulnerabilityComponentsSolutions}, where the relation between the vulnerabilities, components, and proposed prevention measures are presented. In addition to the components exploited from the samples, we have included theoretical target components in the table whenever it is safe to assume that the vulnerability holds for additional components. For instance, we have only seen one case of bad access control located in the debt issuer component. However, this would also be a problem for other smart contract-based components, i.e., if the withdrawal functions in the custodian and liquidity pool were external. We describe each of the 8 identified vulnerabilities throughout the remainder of this section.

\begin{table*}
  \centering
  \caption{The table shows all effective exploits (30) and ~\textsuperscript{1}:  Failed exploits or known bugs (4) found in the timespan of 2021-2023. Each row shows the bridge, the main and eventual secondary vulnerability. The secondary vulnerability indicates some cases with additional layers of complexity. Furthermore, the rows include the main and accessory components. The logic is that the vulnerability is often located elsewhere than the stolen assets. In these cases, the accessory vulnerability is the extraction point.}
  \label{tab:bridgeExploitTarget}
  \scriptsize
  \setlength{\tabcolsep}{2pt}
  \begin{tabular}{|c|c|c|c|c|c|c|}
    \hline
    \textbf{Date} & \textbf{Bridge} & \textbf{Main vulnerability} & \textbf{Secondary vulnerability} & \textbf{Vulnerable component} & \textbf{Accessory component} & \textbf{Root cause} \\
    \hline
    Nov-23 & Heco bridge & Private key leak & - & Attester & Custodian & Unknown \\
    \hline
    Jun-22 & Harmony & Private key leak & - & Attester & Custodian & Unknown \\
    \hline
    Mar-22 & Ronin & Private key leak & - & Attester & Custodian & Phishing \\
    \hline
    Apr-22 & Marvin Inu & Private key leak & - & Attester & Custodian & Unknown \\
    \hline
    Jul-23 & Multichain & Private key leak & - & Attester & Custodian and & Unknown \\
    & & & & & liquidity pool & \\
    \hline
    Jul-21 & Levyathan & Private key leak & - & Attester & Debt issuer & Private key \\
    & & & & & & in GitHub \\
    \hline
    Jun-21 & Bondly & Private key leak & - & Attester & Debt issuer & Unknown\\
    \hline
    Nov-22 & pGala & Private key leak & - & Attester & Debt issuer & Alleged private \\
    & & & & & & key in GitHub \\
    \hline
    Jul-23 & Poly network & Private key leak & - & Attester & Debt issuer & Unknown \\
    \hline
    Oct-22 & QAN Platform & Private key leak & - & Attester (deployer) & Liquidity pool & Low entropy \\
    & & & & & & vanity address \\
    \hline
    Jul-21 & Multichain & Private key leak & - & Attester & Liquidity pool & Same k twice in \\
    & & & & & & ECDSA signatures \\
    \hline
    Nov-22 & Ankr & Private key leak & Privilege escalation & Deployer & Debt issuer & Unknown \\
    \hline
    Jul-22 & BiFi & Private key leak & Privilege escalation & Deployer & Custodian & Unknown \\
    \hline
    Dec-21 & Badger & Private key leak & - & Bridge application & Custodian & Unknown \\
    & & & & backend & & \\
    \hline
    Aug-22 & deBridge~\textsuperscript{1} & Private key leak~\textsuperscript{1} & - & - & - & Phishing \\
    \hline
    Jan-22 & Multichain & Authentication bypass & - & Custodian & Token interface & Non-reverting \\
    & & & & & & fallback \\
    \hline
    Jun-21 & Zapper~\textsuperscript{1} & Authentication bypass & - & Custodian & Token interface & Non-reverting \\
    & & & & & & fallback \\
    \hline
    Aug-21 & Poly Network & Authentication bypass & Privilege escalation & Custodian & - & Forge sighash point \\
    & & & & & & to malicious contract \\
    \hline
    Aug-22 & Nomad & Authentication bypass & - & Custodian & - & Bad contract \\
    & & & & & & initialization \\
    \hline
    Feb-22 & Wormhole & Authentication bypass & - & Debt issuer & - & Assign malicious sig- \\
    & & & & & & verification contract \\
    \hline
    Jul-21 & Chainswap & Authentication bypass & - & Debt issuer & - & Accept Tx without \\
    & & & & & & valid signature \\
    \hline
    May-22 & QAN Platform & Authentication bypass & Bad business logic & Attester & Liquidity pool & No check on Tx value \\
    & & & & & & or deposit contract \\
    \hline
    Jun-22 & Rainbow (Aurora)~\textsuperscript{1} & Fraudulent proof & - & Debt issuer & Custodian & Accepts and signs \\
    & & & & & & fake proof of burn \\
    \hline
    Oct-22 & Binance bridge & Fraudulent proof & - & Custodian & Debt issuer &  Forged IAVL proof\\
    \hline
    Jul-21 & ThorChain & Bad business logic & - & Liquidity pool & Liquidity pool & Does not check \\
    & & & & & & Tx value \\
    \hline
    Jan-22 & Qubit & Bad business logic & - & Custodian & Debt issuer & Does not check \\
    & & & & & & Tx value \\
    \hline
    Feb-22 & Meter.io & Bad business logic & - & Custodian & Debt issuer & Send wrapped tokens \\
    & & & & & & without lock or burn \\
    \hline
    Nov-21 & Zenon & Bad access control & - & Debt issuer & - & Unrestricted access \\
    & & & & & & to burn function \\
    \hline
    Jul-21 & ThorChain & Malicious contract event & Fraudulent memo & Liquidity pool & - & No verification \\
    & & & & & & of event origin \\
    \hline
    Sep-21 & pNetwork & Malicious contract event & - & Debt issuer & - & No verification\\
    & & & & & & of event origin \\
    \hline
    Oct-21 & Polygon Plasma~\textsuperscript{1} & Replay attack & - & Custodian & - & Burn receipt \\
    & & & & & & uniqueness unchecked \\
    \hline
    Sep-22 & Gnosis Omni bridge & Replay attack & - & Custodian & - & Receiver ChainID \\
    & & & & & & uniqueness unchecked \\
    \hline
    May-22 & Rainbow (Near) & Front running & - & Watcher & - & Can frontrun \\
    & & & & & & fraud-proof \\
    \hline
    Aug-22 & Rainbow (Near) & Front running & - & Watcher & - & Can frontrun \\
    & & & & & & fraud-proof \\
    \hline
  \end{tabular}
\end{table*}


\begin{table*}
  \centering
  \caption{An overview of the identified vulnerabilities, the potentially exposed components and prevention measures. ~\textsuperscript{1}: Components are included when it is safe to assume that they are relevant to the vulnerability. These components are likely to be targeted through the corresponding vulnerabilities in the future.}
  \label{tab:vulnerabilityComponentsSolutions}
  \begin{tabular}{|c|c|c|}
       \hline
       \textbf{Vulnerability} & \textbf{Components} & \textbf{Prevention} \\
       \hline
       Private key leak & Attester, deployer, bridge application & Proper multi-signature scheme and  \\
       & frontend~\textsuperscript{1}, backend, and DNS~\textsuperscript{1} & multi-factor authentication. \\
       & custodian, dept issuer, liquidity pool & \\
       \hline
       Authentication bypass & Attester, custodian, debt issuer & Verify the privilege of source and destination addresses of \\
       & token interface & incoming and outgoing smart contract calls. \\
       && Proper error handling. Proper signature verification. \\
       \hline
       Fraudulent proof & Custodian, debt issuer, liquidity pool~\textsuperscript{1} & Strong verification. Verify state. \\
       & & Only the attester can send proofs. \\
       \hline
       Bad business logic & Attester, custodian, debt issuer,  & Ensure that only valid actions can be taken for \\
       & liquidity pool & every logical step in the business process. \\
       \hline
       Bad access control & Attester~\textsuperscript{1}, custodian~\textsuperscript{1}, debt issuer, & Critical functions must be privileged. \\
       & liquidity pool~\textsuperscript{1} & \\
       \hline
       Malicious contract event & Custodian~\textsuperscript{1}, debt issuer, liquidity pool & Verify event origin. An event should come  \\
       && from a trusted source. \\
       \hline
       Replay attack & Custodian, debt issuer~\textsuperscript{1}, liquidity pool~\textsuperscript{1} & Ensure transaction and call data uniqueness. \\
       \hline
       Front running & Watcher & Commit-reveal scheme with zero-knowledge proof. \\
       \hline
  \end{tabular}
\end{table*}

\subsection{Vulnerabilities resulting in private key leak}

Private key leakage is the primary vulnerability for bridges, covering almost half of the exploits in consideration. Key management should be the first concern for bridge operators. The findings demonstrate many single-key and single-entity-operated bridges, which is a high-risk liability.

The private keys are used to engage in different privileged activities, often performed through the attester component. This includes approving and releasing funds from a custodian or pool, minting new tokens from a debt issuer, or burning tokens in a debt issuer. The privileges could also be linked to the deployer, which can change smart contract pointers. This is useful when upgrading a contract where the deployer can route the main bridge contract to a newly deployed upgraded contract. The incidents show an example of privilege escalation~\cite{ankrAttack} where an attacker uses the deployer to point to a malicious contract, allowing them to mint tokens. Another example is setting a deposit address (custodian) controlled by the attacker, meaning they can send funds to their own address without locking before receiving funds on the destination chain~\cite{BiFiAttack}.

Ronin\footnote{\url{https://roninchain.com}} and Multichain\footnote{\url{https://multichain.org/}} are examples of single-entity-operated bridges. In the first case, Sky Mavis\footnote{\url{https://skymavis.com/}} controlled five validators in a scheme requiring five of nine validators to authorize control over the custodian and debt issuer~\cite{roninAttack}. The attacker gained control of the bridge by compromising a single computer operated by Sky Mavis. In Multichain, all the validators were operated from the CEO's Azure account~\cite{multichainAttack-3}. This scheme fell apart after the CEO was arrested, and Multichain is no longer operative.

Private key issues could also be unrelated to blockchain interaction, for instance, a cloud service for the user interface. An example is the Badger bridge\footnote{\url{https://badger.com/}}, where the Cloudflare account for the application was compromised. The attacker made a malicious script that made users approve a malicious contract to control their funds. Similarly, an attacker could get control of the frontend, or the bridge application's DNS configuration to point the users to a malicious website.


Private key leak exploits mainly target the attester, which can perform privileged actions on the bridge. The deployer will also be a target. Otherwise, bridge application components, frontend, backend, and DNS can be exploited to make users interact with malicious applications that send their funds to an adversary. Furthermore, all the private-key leak vulnerabilities target an accessory component, where the assets can be drained. The custodian, debt issuer, and liquidity pools will be the extraction points for private key leaks.

The root causes of private key leaks are often unclear, and whether a leaked private key was caused intentionally for the founders to gain profit (AKA 'rug pull'), or if the key was stolen through phishing can only be speculated without proof. However, the target component and prevention measure will be the same either way.

There have also been cases of sloppy key generation and usage. In one case, the QAN platform used a third-party service and generated their keys with low entropy~\cite{QANAttack-2}, making them vulnerable to reverse key generation based on the public key. Another cryptographic mistake was made by Multichain, who used the same k twice in ECDSA signatures~\cite{multichainAttack-0}.

To mitigate the risks of private key leaks, we recommend that bridges ensure a strong multi-signature scheme to prevent private key leaks. These keys should be isolated; only one key should be available per attester or deployer. Furthermore, keys should be rotated regularly. These prevention measures apply to all the target components that handle private keys: The attester, deployer, backend bridge frontend, backend, and DNS.

It is also recommended to use different schemes for different operations in each related component, for instance, a peg-in, peg-out, and deployment. This prevention measure is relevant for the accessory components that are operated by the private key holders: Custodian, dept issuer, and liquidity pool.

\subsection{Vulnerabilities resulting in authentication bypass}

Authentication bypass also represents many of the bridge exploits. The nature of this vulnerability varies for the different attacks, and we have identified three ways of authentication bypass. The most prominent variant is when the attacker can forward a transaction to a malicious contract. Furthermore, some function calls might fail and fallback without revering. Additionally, there has been a case where null-value signatures have been accepted. Authentication bypass exploits often circumvent the attester completely and relay the transaction directly to a smart contract such as the custodian, debt issuer, or liquidity pool.

Poly network~\cite{polyNetworkAttack}, Wormhole~\cite{wormholeAttack}, and QAN platform~\cite{QANAttack-1} were all victims of attacks where the adversary could assign a malicious contract as part of the authentication process. This allowed them to specify the authentication mechanism to their liking. The Multichain~\cite{multichainAttack-1} and Zapper~\cite{zapperAttack} vulnerabilities were identical, and the attacker could circumvent the permit check by calling the bridge function to deposit assets to a chosen address using {\tt depositWithPermit}. This vulnerability was applicable for tokens whose interfaces did not implement {\tt depositWithPermit}. The token interface would fall back without throwing an error, and the attacker could send users' funds already approved for use by the bridge. In Nomad's case, it was wrongly initialized to accept null (0x00) as a valid proof~\cite{nomadAttack}. Chainswap would accept any signature as long as it was unique.

In summary, the authentication bypass exploits target vulnerable smart contract-based components (custodian, debt issuer, and liquidity pool). Token interface smart contracts are included as accessory components because bridges use those to access users' funds residing in their wallets. However, bridge developers often do not own and cannot change token interfaces and must handle them as a vulnerability in their own bridge components. Therefore, the bridge smart contracts must be responsible for reverting an invalid transaction instead of relying on external modules. 

Some prevention measures involve the attester in addition to the mentioned smart contract-based components. These measures can be done by verifying the authentication for incoming and outgoing contract calls in the bridging process. At last, signatures and proofs must be adequately verified. The related proof and signature verification vulnerabilities were considered authentication bypasses because validator checks could be bypassed rather than forging proof or signature.





\subsection{Vulnerabilities resulting in privilege escalation}

Privilege escalation can happen when it is possible to assign new contracts or addresses that are part of privileged processes. The root causes are related to private key leaks and authentication bypasses, and so are the preventive measures. This is why we do not elaborate on prevention measures or list this vulnerability and discuss its exploitation impact in~\cref{tab:extractionPointImpact}.

BiFi~\cite{BiFiAttack} is an example of an attacker getting hold of a privileged key to assign deposit addresses. The attacker set their own address as a deposit address. The bridge would record the deposits as valid and mint assets from the debt issuer as the accessory component. The attacker could then freely control the funds in the deposit address and the minted assets.
\subsection{Vulnerabilities resulting in fraudulent proof}

An attacker can exploit bridges by forging fraudulent proofs. The vulnerable component depends both on where proofs are generated and where they are verified. The relevant events are proofs of deposit (custodian \& liquidity pool) and proofs of burn (debt issuer).

The bounty hunter who found a bug on the Aurora side of the Rainbow Bridge received a 1M\$ reward. The issue was that an adversary could specify the amount, receiver address, and custodian address and pass it through Aurora's {\tt view} function~\cite{auroraBug}, which makes it seem like the debt issuer vetted and executed the burn transaction. In turn, the specified amount could be withdrawn from the custodian.

The samples on fraudulent proof show that the vulnerable component and the accessory component (the extraction point) often reside on different sides of the victim bridge. The relevant components are the Custodian, debt issuer, and liquidity pool because they generate or receive the proofs.


State and execution validation can be used in the custodian, debt issuer, and liquidity pool to abort upon invalid execution or state change and prevent fraudulent proof exploits from extracting any value. It could also be advantageous if these components only allow proofs from trusted sources.

\subsection{Vulnerabilities resulting in bad business logic}


Bad business logic vulnerabilities occur because the software has been implemented incorrectly and allows invalid actions. An adversary may inspect the range of valid input variables and attempt to skip steps in the business process. For instance, one could deposit a specified amount without attaching any value.

In both ThorChain~\cite{thorChainAttack} and Qubit~\cite{qubitAttack}, the bridge would accept a deposit with zero value attached. Instead, the value was specified as a parameter, and this was the number checked by the contracts. In another case, Meter~\cite{meterAttack}, the attacker could deposit wrapped tokens without locking, skipping this logical step in the business process. Afterwards, assets could be withdrawn on the source and destination chains.

The prevention measures for bad business logic must be implemented in the attester, custodian, debt issuer, and liquidity pool. The measures are to verify the logical operations in the business process continuously. For instance, it should be verified that assets are locked, minted or burned. Another preventive measure for these components is to make an accounting mechanism that compares the state of locked and synthetic tokens to ensure they are equal.

\subsection{Vulnerabilities leading to bad access control}

Some critical functions should only be callable by privileged or verified entities. For instance, withdrawal from the custodian, liquidity pool, or mint and burn in the debt issuer. By mistake, the Zenon Network~\cite{zenonAttack} declared the burn function external (i.e., public) and allowed the attacker to burn tokens they did not own. The attacker bought into a position in a corresponding DEX before burning other tokens to pump the value of their own tokens.

Bad access control vulnerabilities should be mitigated by restricting access to privileged functionality in the attester, custodian, debt issuer, or liquidity pool.

\subsection{Vulnerabilities resulting in malicious contract event}

Bridges often use on-chain events to propagate cross-chain transfers. However, these events are not unique or distinguished by origin, and an adversary can create a contract that emits events in the same format as the bridge contract. This happened in the exploits of ThorChain~\cite{thorChainAttack-2} and pNetwork~\cite{pNetworkAttack}. 

The relevant components emit events upon actions such as deposit or burn. These actions are performed by smart contract-based components: the custodian, debt issuer, and liquidity pool. These components should include additional data in events to prove and identify events coming from authorized sources and avoid malicious contract events.

\subsection{Vulnerabilities resulting in replay attack compromises}

Replay attacks happen when it is possible to rebroadcast a transaction on another chain or create a new transaction reusing the same call data as the original. For Polygon Plasma~\cite{polygonPlasmaAttack} and Gnosis Omni Bridge~\cite{gnosisAttack} the replays were executed against the custodian which would accept the same deposit transaction or call data twice before issuing assets on the destination chain. This vulnerability could also affect the debt issuer or liquidity pool. For instance, the debt issuer would be affected if a mint transaction can be submitted twice.

To prevent replay attacks, bridges must verify the uniqueness of transactions and call data whenever a contract call is made toward the custodian, debt issuer, or liquidity pool.


\subsection{Vulnerabilities resulting in front running compromises}

Front running and other maximum extractable value (MEV) attacks are most relevant for inter-chain exchanges. However, it can become a problem for the watchers securing optimistic bridges. In the Rainbow Bridge~\cite{rainbowAttack} MEV bots were able to intercept, and front run fraud proofs to reap the rewards.
At first, this can be seen as theft from the watcher. Additionally, it might jeopardize the incentive mechanisms for optimistic bridges, undermining the security assumption of having active, trusted watchers.

We suggest using a commit-reveal scheme combined with zero-knowledge proofs. The commit transaction does not reveal any useful information to bots and has no value attached to it. In a subsequent block, the watcher can reveal the zero-knowledge proof to prove the validity of the fraud-proof without revealing its content. This is one of the approaches being discussed in the Ethereum foundation to reduce the impact of MEV bots~\cite{MEVBots}. 

In bridges, MEV bots are only known to frontrun watchers. However, for other DeFi applications, such as DEXes, there are also liquidity pools targeted by MEV bots. These pools operate on the same chain and are more susceptible to the bots. The exchange is harder to detect in bridges, and an MEV bot might be unable to place the buy-and-sell order for a sandwich attack on the same block because they reside in different chains. 

Recent research has been initiated for cross-domain MEV~\cite{crossDomain}. However, the documented attacks require bridge aggregators using swaps, as shown in~\cref{fig:aggregator}. Since we strictly limit ourselves to bridging, excluding aggregation and swaps, these exploits are out of scope. On the other hand, we assume that some arbitrage vulnerabilities in liquidity network bridges could also occur. To take this notion, we have included price calculation as a component for bridge architectures (\cref{tab:componentsArchitectures}).



\section{RQ3: Impact reduction}
\label{sec:impactReduction}

Throughout our study, we saw several cases of bridges attempting to reduce the impact of exploits as soon as they were discovered. Although preventive measures should be the first priority, it is also reassuring that there are plans in case a crisis is imminent. In this section, we will discuss the details of such damage reduction measures listed in~\cref{tab:extractionPointImpact}. The figure depicts how different entities have played a part in reducing the impact of exploits. Some impact reduction measures may require centralised or external intervention, which is often frowned upon in the blockchain community. However, for bridges, these measures could be critical and sometimes a necessary evil required for their survival.

Note that the impact reduction measures in the table relate to almost all the same target components for every single measure. That is because impact reduction measures target the extraction point as seen in~\cref{fig:image2}. As described before, the extraction point might be the accessory component, which is not where vulnerabilities reside. Therefore it is not viable to make connections between vulnerabilities and impact reduction measures. However, we can state that the measures can be enabled for all the smart contract-based components (custodian, debt issuer, liquidity pool, and token interface). An exception might be the watcher if you can outbid another MEV bot.

\begin{table*}
  \centering
  \caption{Shows how a responsible entity can react when certain components are exploited to engage impact reduction measures.}
  \label{tab:extractionPointImpact}
  \begin{tabular}{|c|c|c|}
       \hline
       \textbf{Entity} & Exploited target component & \textbf{Impact reduction measure} \\
       \hline
       Bridge operators & Custodian, debt issuer, liquidity pool \& token interface & Higher security for large transfers \\
       \hline
       & Custodian, debt issuer, liquidity pool \& token interface &  Max transfer size \\
       \hline
       & Custodian, debt issuer, liquidity pool \& token interface & Increase the fraud window size \\
       \hline
       & Custodian, debt issuer, liquidity pool \& token interface &  Pause withdrawal, deposit, mint and burn \\
       \hline
       & Custodian, debt issuer, liquidity pool \& token interface &  Allocate resources and staff to monitor and react \\
       \hline
       &  Custodian, debt issuer, liquidity pool, token interface, watcher & Monitoring and automatic recovery \\
       \hline
       Blockchain validators &  Custodian, debt issuer, liquidity pool \& token interface & Pause the blockchain \\
       \hline
       & Custodian, debt issuer, liquidity pool \& token interface &  Freeze or seize assets \\
       \hline
       Exchange operators & Custodian, debt issuer, liquidity pool \& token interface &  Freeze or seize assets \\
       \hline
       Token operators &  Custodian, debt issuer, liquidity pool \& token interface & Freeze or seize assets \\ 
       \hline
       Law enforcement &  Custodian, debt issuer, liquidity pool \& token interface & Blacklist funds \\
       \hline
  \end{tabular}
\end{table*}

\subsection{Bridge operators}

\subsubsection{Higher security for large transfers}

Bridges could implement \textit{higher security for large transfers}. For instance, delaying large transactions allows for intervention. Another approach is to increase the threshold for multi-signature schemes for large transactions.

\subsubsection{Max transfer size}
Similarly, bridges can set a~\textit{max transfer size}. The limit can increase depending on the maturity of the bridge or for individual users depending on their trust score. These measurements might not affect most users who make modest transfers across bridges. Most of the exploits seen in this study are in the millions of dollars. Bridges may drastically reduce the impact of these exploits by establishing transfer limits.

\subsubsection{Increase the fraud window size}

Optimistic bridges may \textit{increase the fraud window size}. This gives watchers more time to detect fraudulent transactions.

\subsubsection{Pause withdrawal, deposit, mint and burn}

Furthermore, bridges should implement capabilities to \textit{pause withdrawal, deposit, mint, and burn}. There should be a plan to decide how and when these emergency mechanisms should be utilised. These plans should be presented to users to provide assurance in case of a security breach.

\subsubsection{Allocate resources and staff to monitor and react}

There have been cases where internal and external developers and hackers have noticed a bug before or during an exploit. For instance, during the Nomad Bridge attack, white hat hackers saw the exploit and started a~\textit{white hat rescue} by attacking the bridge themselves to secure and return funds before they could be stolen by the attackers~\cite{nomadAttack}. It could be a viable choice to hack your own bridge when you are out of options. For this measure, we recommend operators assign resources to monitor and an available task force to address and counter the attacks. 

\subsubsection{Monitoring and automatic recovery}

MEV bots frontrun lucrative transactions regardless of what the underlying purpose might be. In some cases, this has enabled accidental rescue of funds by frontrunning an attacker's exploit. This happened during the Nomad exploit, where the bot operator committed to return the stolen funds that they seized by frontrunning a hacker~\cite{nomadAttack}. In other words, a bridge could set up MEV bots to monitor and target their own smart contracts to automatically detect, frontrun, and recover attacks. This measure would also increase the resilience of a bridge or corresponding swap toward arbitrage attacks.

\subsection{Blockchain validators}

Some blockchains are criticised for being too centralised and that their system could become subject to external control, asset seizures, or transaction censorship. However, for bridges, these attributes can be critical for impact reduction. The underlying blockchain might be able to~\textit{pause the blockchain} and ~\textit{freeze or seize} assets after they have left the bridge's possession. This happened for the Binance bridge exploit, where the debt issuer was compromised and the attacker started minting tokens. The validators reacted by pausing the blockchain and froze assets worth around 430 million dollars and reduced the impact of the exploit, which was initially able to drain 586 million dollars worth of assets from the bridge~\cite{bnbAttack}.

These measurements require substantial influence over the underlying blockchain, which many bridge operators do not have. This measurement is most viable for bridges such as Binance's, which also operate the underlying blockchain. Other bridges might have to establish contact with consensus validators prior to an attack as a preparatory measure.

\subsubsection{Pause the blockchain}
If the blockchain stops processing blocks then the attacker cannot bridge funds to another chain or send the assets into a mixer, such as Tornado Cash~\cite{tornadoCash}.

\subsubsection{Freeze or seize assets}
Make the assets unavailable to the attacker or return them to the original owner.


\subsection{Exchange operators}

\subsubsection{Freeze or seize assets}

Exchanges could play a vital role in stopping hackers from transporting stolen funds. This could be especially useful if the attacker is able to mint tokens "out of thin air". Exchanges could then~\textit{freeze assets} to prevent the token value from being affected until the situation has been resolved. An example of this method can be seen from the pGala exploit~\cite{pGalaAttack}. A centralised exchange might also seize stolen assets if the attacker tries to funnel them through the exchange.

To implement this measure one must either control corresponding exchanges or establish connections to exchange operators to prepare for a worst-case scenario.

\subsection{Token operators}

\subsubsection{Freeze or seize assets}

Some tokens are designed to be more centralised than others and may have the capability to~\textit{freeze or seize assets}. For instance, after the Poly Network exploit, Tether decided to freeze 33 million USDT on the Ethereum chain to prevent the attacker from getting away with those assets~\cite{polyNetworkAttack}. This measure would also require close contact with the corresponding token operators.

\subsection{Law enforcement}

\subsubsection{Blacklist funds}

The USA government took action after the North Korean Lazarus group was identified as the suspected adversary in the exploits of Ronin~\cite{roninAttack} and Harmony~\cite{harmonyAttack}. The Office of Foreign Assets Control (OFAC) has decided to~\textit{blacklist funds}~\cite{sanctions} that have gone through the Tornado cash mixer\footnote{\url{https://github.com/tornadocash}}. These measurements could prevent hackers from utilizing stolen funds.

This impact reduction measure is not relevant for bridge operators because sanctions do not help them recover lost value. However, it could be a viable measure to limit the impact of DeFi exploits. 


\section{Discussion}
\label{sec:discussion}

\subsection{Comparison with related work}
\label{sec:relatedWork}

Lee et al.~\cite{10174993} present a systemisation of knowledge on bridge exploits. The scope is similar as we both focus solely on bridges. The article summarises some security incidents and presents them as individual vulnerabilities. The authors worked with an abstracted bridge architecture that assumes the presence of a custodian, communicator, and debt issuer. Our findings do however show that an abstracted architecture does not work well to differentiate between the nuances of cross-chain bridges. Most of the nine vulnerabilities included in ~\cite{10174993} are covered by our study. However, we claim our novelty not only by summarising vulnerabilities but also by assessing architectural relations to vulnerabilities, prevention and impact reduction measures.

Zhang et al.~\cite{10.1145/3551349.3559520} proposed three bug classes for cross-chain bridges and Xscope as a bridge monitoring system to detect bridge attacks. The assessed attacks partly overlap with our analysis. In summary, Xscope is an attack detection tool which validates the execution of cross-chain transactions by monitoring smart contract events. In contrast to focus on attack detection, we explore cross-chain vulnerabilities and their relation to cross-chain architectures. After seeing the nuances of cross-chain bridges we would be curious to understand how well Xscope can work for different bridges and whether it applies on a general level, work for some bridge architectures and attack, or has to be tailored for individual architectures and exploits.

Li et al.~\cite{9881607} made a security analysis of vulnerabilities, attacks and optimisations in Defi. The article has a wide coverage for smart contracts and DEXes. On the other hand, it includes few bridge samples, and most attacks are out of scope. The main similarity to this article is that it also emphasises private-key leaks as a prominent vulnerability. Other bridge-related vulnerabilities were grouped as individual "other bugs". In comparison, we have focused solely on bridges to understand the architectures and link them to vulnerability categorization and prevention measures.










\subsection{Implications}


Bridge startups are often small teams, and there is a race to get into the market and be competitive. It seems like these conditions might have encouraged some of the reckless bridge architectures that have enabled many exploits. Even though there are many good practices, security has not been taken seriously, and we see this for many bridges. For instance, a solid multi-signature scheme has been a known bridge requirement for a long time. Still, developers are taking shortcuts and often operate with just one or a few attester keys. Although key management and security of private keys should be a top priority, we still see keys being exposed openly on GitHub, and many bridges do not clearly communicate how attesters are managed.

Throughout this study, we have systematised knowledge about cross-chain architectures, components, exploits, and their relation to vulnerabilities and countermeasures. We believe these insights can be valuable and directly applicable for bridge developers to secure their assets. Our results can be seen as a checklist (Tables~\ref{tab:componentsArchitectures},~\ref{tab:vulnerabilityComponentsSolutions}, and~\ref{tab:extractionPointImpact})
for developers to take preventive measures for the components in their specific architecture and prepare for damage reduction. Here, we provide one example to demonstrate how the findings from this study can be used as a checklist when designing a bridge: First, one should select the bridge architecture that has the most desirable attributes corresponding to the needs. In this case, I chose the optimistic token bridge architecture because it can be designed without worrying about attesters and has a weak trust assumption. I made a token bridge because it is easier to on-ramp as it requires no liquidity in pools. If I choose to implement only the mandatory components, then I need to focus on the relayer, deployer, watcher and liquidity pool components. Since all of the vulnerabilities are related to at least one of the components in this architecture, we would have to assess them individually and apply countermeasures as proposed in~\cref{tab:vulnerabilityComponentsSolutions}. All the impact reduction measures could be relevant to the bridge I want to design. A simple starter would be to set a low max limit on the bridge until I can see that it is working. I would also like to monitor the activity and enable functionality to pause the bridge if something goes wrong.




\section{Conclusion and future work}
\label{sec:conclusion}

This article presents 13 critical components for blockchain bridges and links them to different bridge architectures. We analysed 34 security incidents in cross-chain bridges to identify eight vulnerability categories that are critical for bridges. Furthermore, the connection between the components and vulnerabilities was highlighted before recommending measures to prevent and reduce the impact of bridge exploits.

The industry and community surrounding bridges are still immature, and we expect more multi-million dollar exploits in the future. However, we hope many incidents could have been avoided by following the knowledge summarised in this study. 

For future work in cross-chain bridging, we believe it will be interesting to follow the adoption of bridges that are more advanced, resilient and decentralized compared to third-party attestation protocols, which are most popular, probably because of their simplicity. 

We also deem the topic of cross-domain MEV as interesting and something that may cause the same amount of disruption for bridges as MEV has already done for DEXes. This topic has many unanswered questions and is an open field for researchers.





\printbibliography

\end{document}